\documentstyle[twocolumn,aps]{revtex}

\begin{document}
\title{Bell's inequalities for states with positive partial transpose}
\author{R.~F. Werner and M.~M. Wolf}
\address{
Institut f\"ur Mathematische Physik, TU Braunschweig,
Mendelssohnstr.3, 38106 Braunschweig, Germany. \\ Electronic Mail:
r.werner@tu-bs.de, mm.wolf@tu-bs.de}

\date{\today}

\maketitle

\begin{abstract} We study violations of $n$ particle Bell
inequalities (as developed by Mermin and Klyshko) under the
assumption that suitable partial transposes of the density
operator are positive. If all transposes with respect to a
partition of the system into $p$ subsystems are positive, the best
upper bound on the violation is $2^{(n-p)/2}$. In particular, if
the partial transposes with respect to all subsystems are
positive, the inequalities are satisfied. This is supporting
evidence for a recent conjecture by Peres that positivity of
partial transposes could be equivalent to existence of local
classical models.
\end{abstract}
\section{Introduction}

One of the basic questions asked early in the development of
quantum information theory was about the nature of entanglement.
Extreme cases were always clear enough: a 2 qubit singlet state
was the paradigm of entangled state \cite{Bohm}, whereas product
states and mixtures thereof were obviously not, but merely
``classically correlated'' \cite{Wer}. But in the wide range
between it was hardly clear where a meaningful boundary between
the entangled and the non-entangled could be drawn. Still today
some boundaries are not completely known, although, of course,
general structural knowledge about entanglement has increased
dramatically in the last few years. The present paper is devoted
to settling the relationship between two entanglement properties
discussed in the literature.

To fix ideas we will start by recalling some properties one might
identify with ``entanglement''  and the known relations between
them. For simplicity in this introduction we will choose the
setting of bipartite quantum systems, i.e., quantum systems whose
Hilbert space is written as a tensor product ${\cal H}={\cal
H}_1\otimes{\cal H}_2$. Moreover, we consider finite dimensional
spaces only, leaving the appropriate extensions to infinite
dimensions to the reader. All properties listed refer to a density
matrix $\rho$ on this space. It turns out to be simpler to define
the entanglement properties in terms of there negations, i.e., the
various degrees of ``classicalness'':

\begin{description}
\item{(S)}  A state is called {\bf separable} or ``classically
correlated'', if it can be written as a convex combination of
tensor product states. Otherwise, it is simply called
``entangled''.

\item{(B)} Before 1990 perhaps the only mathematically sharp
criterion for entanglement were the Bell inequalities in their
CHSH form \cite{chsh}. A state is said to satisfy these {\bf Bell
inequalities} if, for any choice of operators $A_i,A_i'$ on ${\cal
H}_i$ ($i=1,2$) with $-{\bf1}\leq A_i,A_i'\leq{\bf1}$, we have
\begin{equation}\label{CHSH}
  {\rm tr}\rho\Big(A_1\otimes(A_2+A_2')+A_1'\otimes(A_2-A_2'
  )\Big)\leq2.
\end{equation}
It is easy to see that (S)\hbox{$\Rightarrow$}(B).

\item{(M)} Bell's inequalities are traditionally derived from
an assumption about the existence of local hidden variables. The
same assumptions lead to an infinite hierarchy of correlation
inequalities \cite{Pit}, and it seems natural to base a notion of
entanglement not on an arbitrary choice of inequality (e.g. CHSH)
from this hierarchy. So we say that $\rho$ {\bf admits a local
classical model}, if it satisfies all inequalities from this
hierarchy. Then (S)\hbox{$\Rightarrow$}(M)\hbox{$\Rightarrow$}(B).
It was shown in \cite{Wer} that (M)\hbox{$\not\Rightarrow$}(S),
and this was perhaps the first indication that different types of
entanglement might have to be distinguished.

\item{(U)} A key step for the development of entanglement
theory was a paper by Popescu \cite{Pop}, showing that by suitable
local filtering operations applied to maybe several copies of a
given $\rho$, one could sometimes obtain a new state $\rho'$
violating a Bell inequality, even though $\rho$ admitted a local
hidden variable model, and hence satisfied the full hierarchy of
Bell inequalities. Let us call a state {\bf undistillible}, if it
is impossible to obtain from it a two qubit state violating the
CHSH inequality, by any process of local quantum operations (i.e.,
operations acting only on one subsystem), perhaps allowing
classical communication, and several copies of the state as an
input. What Popescu showed was that
(M)\hbox{$\not\Rightarrow$}(U).

\item{(P)} The idea of distillation was later taken to much
greater sophistication \cite{dis1}, and for a while the natural
conjecture seemed to be that not only (S)\hbox{$\Rightarrow$}(U)
(which is trivial to see), but that these two should be
equivalent. The counterexample was provided in \cite{Hor}. They
used a property (P), which had  been proposed by Peres \cite{Per}
as a necessary condition for separability (i.e.,
(S)\hbox{$\Rightarrow$}(P)), which turned out to be also
sufficient in the qubit case \cite{Wor}. This condition (P) is
that $\rho$ has {\bf positive partial transpose}, i.e.,
$\rho^{T_1}$ is a positive semi-definite operator. Here the
partial transpose $A^{T_1}$ of an operator $A$ on ${\cal H}={\cal
H}_1\otimes{\cal H}_2$ is defined in terms of matrix elements with
respect to some basis by
 \begin{equation}\label{pt}
  \langle k\ell\vert A^{T_1} \vert mn \rangle
   = \langle m\ell\vert A\vert kn \rangle.
\end{equation}
Equivalently,
\begin{equation}\label{pt2}
  (\sum_\alpha A_\alpha\otimes B_\alpha)^{T_1}
      = \sum_\alpha A_\alpha^T\otimes B_\alpha,
\end{equation}
where the superscript $T$ stands for transposition in the given
basis. It was shown that (P)\hbox{$\Rightarrow$}(U), and the
counterexample in \cite{Hor} worked by establishing (U) and
not-(S) in this example. States of this kind are now called {\it
bound entangled}.
\end{description}
There are further interesting properties, like usefulness for
teleportation \cite{Pop2}, but the above are sufficient for
explaining the problem addressed in this paper. To summarize, it
is known that (S)\hbox{$\Rightarrow$}(P)\hbox{$\Rightarrow$}(U),
and (S)\hbox{$\Rightarrow$}(M)\hbox{$\Rightarrow$}(B). For pure
states all conditions are equivalent, and for systems of two
qubits (U)\hbox{$\Rightarrow$}(S), but
(M)\hbox{$\not\Rightarrow$}(S).

For multi-partite systems, i.e., systems with Hilbert space ${\cal
H}_1\otimes{\cal H}_2\otimes\cdots\otimes{\cal H}_n$, the
properties (S),(M),(U) immediately make sense. For (B) there may
be several choices of inequalities following from (M). The
inequalities we use in this paper are discussed in detail in the
next section. Partial transposition (P) is an intrinsically
bipartite concept. The strongest version of (P) in multi-partite
systems is the one we use below: the positivity of partial
transposes with respect to every subsystem.

Then the implication chains
(S)\hbox{$\Rightarrow$}(P)\hbox{$\Rightarrow$}(U), and
(S)\hbox{$\Rightarrow$}(M)\hbox{$\Rightarrow$}(B) hold as in the
bipartite case. However, no direct relations were known so far
between these chains, even in the bipartite case. It seems likely
that the violation of (B) is a fairly strong property, perhaps
implying distillibility. This certainly seems to be the intuition
of Peres in \cite{Per2} who conjectures that
\begin{equation}\label{conject}
  {\rm (M)} \Longleftrightarrow {\rm(P)}.
\end{equation}
 We will refer to this statement as {\bf Peres' Conjecture}.
It should be noted, however, that neither we nor Peres gave a
sharp mathematical formulation, particularly of the way the model
is required to cover not only one pair but also tensor products
and distillation processes. Some such condition is certainly
needed (and implicitly assumed by Peres), because otherwise the
implication ${\rm (M)} \Rightarrow {\rm(P)}$ would fail already
for two qubits \cite{Wer}. It is not entirely clear from
\cite{Per2} how strongly Peres is committed to (\ref{conject}).
Personally, we would not place a large bet on it. However, we do
follow Peres' lead in seeing here an interesting line of inquiry.
Indeed, the present paper is devoted to proving one special
instance of the conjecture, namely the implication
(P)\hbox{$\Rightarrow$}(B), for general multi-partite systems,
where (P) is taken as the positivity of {\it every} partial
transpose, and (B) is taken as the $n$-particle generalization of
the CHSH inequality proposed by Mermin \cite{Mermin}, and further
developed by Ardehali \cite{Ardehali}, Belinskii and Klyshko
\cite{Klyshko} and others \cite{Roy,Gisin}.

\section{Mermin's generalization of the CHSH inequalities}

Like the CHSH inequalities, Mermin's $n$-party generalization
refers to correlation experiments, in which each of the parties is
given one subsystem of a larger system, and has the choice of two
$\pm1$-valued observables to be measured on it. The expectations
of such an observable are given in quantum mechanics by a
Hermitian operator $A$ with spectrum in $\lbrack-1,1\rbrack$, and
with a choice of $A_k,A_k'$ at site $k$ the raw experimental data
are the $2^n$ expectation values of the form
 ${\rm tr}(\rho\,A_1\otimes A_2'\otimes\cdots A_n)$ with all possible
choices $A_k$ vs. $A'_k$ at all the sites.

If we look only at a single site, the possible pairs of
expectation values (with fixed $A,A'$ but variable $\rho$) lie in
a square. It will be very useful for the construction of the
inequalities and the proof of our result to consider this square
as a set in the complex plane: after a suitable linear
transformation (a $\pi/4$-rotation and a dilation) we can take it
as the square ${\cal S}$ with the corners $\pm1$ and $\pm i$. The
pair of expectation values of $A$ and $A'$ is thus replaced by the
single complex number ${\rm tr}(\rho a)$, where
 \begin{eqnarray}
    a&=&{1\over2}\ \bigl((A+A')+i(A'-A)\bigr)
\label{a-def}\\
     &=&e^{-i\pi/4}\,(A+iA')/\sqrt2.
\label{a-rot}%
\end{eqnarray}
 The idea of this transformation is that the square
${\cal S}$ has a special property: products of complex numbers
$z_k\in{\cal S}$ lie again in ${\cal S}$. This is evident for the
corners (they form a group) and follows for the full square by
convex combination. Suppose now that
$\rho=\bigotimes_{k=1}^n\rho_k$ is a product state. Then the
operator $b=\bigotimes_{k=1}^na_k$ has expectation ${\rm
tr}(\rho\,b)=\prod_{k=1}^n{\rm tr}(\rho_k a_k)\in{\cal S}$. Since
the expectation is linear in $\rho$, the same follows for any
separable state, i.e., any convex combination of product states.
The statement ``${\rm tr}(\rho\,b)\in{\cal S}$'' is essentially
{\bf Mermin's inequality}, although not yet written as an
inequality. Note that the argument given here implies also that
this statement (written out in correlation expressions involving
$A_k,A_k'$) holds in any local classical model, because in a
classical theory every pure state of a composite system is
automatically a product, hence every state is separable. Thus
Mermin's inequality indeed belongs to the broad category of Bell's
inequalities.

To write ``${\rm tr}(\rho\,b)\in{\cal S}$'' as a bona fide set of
inequalities, we just have to undo the
transformation~(\ref{a-def}), i.e., we introduce operators $B,B'$
such that (\ref{a-def}) is satisfied with $(b,B,B')$ substituted
for $(a,A,A')$. The operators $B,B'$ are usually called {\bf Bell
operators}, and Mermin's inequality simply becomes
\begin{equation}\label{mermin}
  \vert{\rm tr}(\rho\, B)\vert\leq1
  \quad\hbox{resp.}\quad
  \vert{\rm tr}(\rho\, B')\vert\leq1.
\end{equation}
 Writing out $B$ and $B'$ explicitly in terms of tensor products of
$A_k,A_k'$ gives the usual CHSH inequality (\ref{CHSH}) for $n=2$,
and becomes arbitrarily cumbersome for large $n$. It is also not
helpful for our purpose. The above derivation also gets rid of the
case distinction ``$n$ odd/even'', which has troubled the early
derivations. In fact, Mermin \cite{Mermin} first missed a factor
$\sqrt2$ for even $n$, which was later obtained by Ardehali
\cite{Ardehali} who in turn missed the same factor for odd $n$.
Inequalities equally sharp for even and odd $n$ were established
in \cite{Roy} and in \cite{Klyshko}.

\section{Violations of Mermin's inequality in Quantum Mechanics}

The idea of combining $A,A'$ in the non-hermitian operator $a$ has
a long tradition for the CHSH case \cite{chshtrad}. Its power is
not only in organizing the inequalities (only linear
transformations among operators are needed for that purpose), but
in the possibility of bringing in the non-commutative algebraic
structure of quantum mechanics to analyze the possibility of
violations in the quantum case. In this section we discuss these
violations, at the same time building up the machinery needed in
the proof of our result. We will need the following expressions:
\begin{eqnarray}
    a^*a&=& \frac12\ (A^2+A'^2)+\frac i2\lbrack{A,A'}\rbrack
\label{a*a}\\
    aa^*&=& \frac12\ (A^2+A'^2)-\frac i2\lbrack{A,A'}\rbrack
\label{aa*}\\
     a^2-a^{*2}&=&i(A'^2-A^2)
\label{aa}%
\end{eqnarray}
 It is clear from the first line that although
${\rm tr}(\rho\,a)$ lies in ${\cal S}$, and hence in the unit
circle for all $\rho$, the operator norm
 $\Vert a\Vert=\Vert a^*a\Vert^{1/2}$ may be $>1$. Therefore, the
tensor product operator $b$ may have a norm increasing
exponentially with $n$. This is the key to the quantum violations
of Mermin's inequality.

The largest possible commutators, i.e., operators saturating the
obvious bound
 $\Vert\lbrack A,A'\rbrack\Vert\leq2\Vert A\Vert\cdot\Vert A'\Vert$
are just Pauli matrices. A good choice is
$A_k=(\sigma_x+\sigma_y)/\sqrt2$ and
$A_k'=(\sigma_x-\sigma_y)/\sqrt2$ for all $k$. Then
$a_k=\sqrt2\,v$, where $v=\pmatrix{0&0\cr1&0}$. It is readily
verified that $v^{\otimes n}$ acts in the two-dimensional space
spanned by $e_1^{\otimes n}$ and $e_2^{\otimes n}$ exactly as $v$
acts in the space spanned by the two basis vectors $e_1,e_2\in{\bf
C}^2$. With the same identification of two-dimensional subspaces
$b=2^{n/2}v^{\otimes n}$ acts like $2^{(n-1)/2}a$, so the possible
expectations ${\rm tr}(\rho\,b)$, with $\rho$ supported in this
subspace span the exponentially enlarged square
 $2^{(n-1)/2}{\cal S}$.

In order to show that $2^{(n-1)/2}$ is the maximal possible
violation (in analogy to Cirel'son's \cite{Cirelson} bound for the
CHSH inequality), but also in preparation of the proof of our main
result it is useful to consider the following general technique
for getting upper bounds on ${\rm tr}(\rho\,b)$. It has been used
in the CHSH case by Landau \cite{Landau}, among others. Note first
that ${\rm tr}(\rho B)$ and ${\rm tr}(\rho B')$ are affine
functionals of each $A_k$ or $A'_k$. Hence if we maximize the
expectations of Bell operators by varying some $A_k$ or $A_k'$,
keeping $\rho$ fixed, we may as well take $A_k$ extremal in the
convex set of Hermitian operators with $-{\bf1}\leq
A_k\leq{\bf1}$. That is to say, we may assume
$A_k^2=A_k'^2={\bf1}$ for all $k$. Taking tensor products of
Equation~(\ref{a*a}) and expanding the product we find
\begin{eqnarray}
  b^*b&=&\bigotimes_{k=1}^n
         \Bigl({\bf1}+ \frac i2\lbrack{A_k,A_k'}\rbrack\Bigr)
\nonumber\\
      &=&\sum_{\beta} \bigotimes_{k\in \beta}
      {i\over2}\lbrack{A_k,A_k'}\rbrack,  \label{b*b}
\end{eqnarray}
where the sum is over all subsets $\beta\subset\{1,\ldots ,n\}$,
and only factors different from ${\bf1}$ are written in the tensor
product. In particular, the term for $\beta=\emptyset$ is
${\bf1}$. For $bb^*$ we get a similar sum with an additional
factor $(-1)^{\vert\beta\vert}$, where $\vert\beta\vert$ denotes
the cardinality of the set $\beta$. From Equation~(\ref{aa}) we
find $a_k^2=a_k^{*2}$, and $b^2=b^{*2}$, by taking tensor
products. Again by applying (\ref{aa}), to $(b,B,B')$ this time,
we find that $B^2=B'^2$. In fact, by adding Equations (\ref{a*a})
and (\ref{aa*}) and inserting Equation~(\ref{b*b}) we get
\begin{eqnarray}\label{Bsq}
  B^2=B'^2&=& {1\over2} (b^*b + bb^*)
\nonumber\\
     &=& \sum_{\beta\ \rm even}\
         \bigotimes_{k\in \beta} {i\over2}\lbrack{A_k,A_k'}\rbrack.
\end{eqnarray}
By the variance inequality $\vert{\rm tr}(\rho B)\vert^2\leq{\rm
tr}(\rho B^2)$ the expectation of the right hand side is an upper
bound on the square of largest violation of Mermin's inequality.
There are two immediate applications: since each term in the sum
has norm at most one, the norm of the sum is bounded by the number
of terms, i.e., $2^{n-1}$. This shows the analog of Cirel'son's
inequality, i.e., that the violation discussed above is indeed
maximal. The second application is to the case that all
commutators vanish. Then only the term for $\beta=\emptyset$
survives, and there is no violation of the inequality. Our result
to be stated and proved in the next section is a refinement of
this idea.

\section{Positive Partial Transposes and Main Result}
We now apply the technique of the previous section to the partial
transpose. More specifically, for any density operator $\rho$ and
any subset $\alpha\subset\{1,\ldots ,n\}$, let $\rho^{T_\alpha}$
denote the partial transpose of all sites belonging to $\alpha$.
Suppose now that $\rho^{T_\alpha}$ is positive semi-definite, and
hence again a density matrix. Then we can apply the variance
inequality to $\rho^{T_\alpha}$ and $B^{T_\alpha}$, obtaining:
\begin{eqnarray}
 ({\rm tr}\rho B)^2
     &=&\bigl({\rm tr}\rho^{T_\alpha} B^{T_\alpha}\bigr)^2
      \leq{\rm tr}\bigl(\rho^{T_\alpha} (B^{T_\alpha})^2\bigr)
 \nonumber\\
     &\leq&{\rm tr}\Bigl(\rho \bigl((B^{T_\alpha})^2\bigr)^{T_\alpha}\Bigr)
\label{var}
\end{eqnarray}
 We note that $\left\lbrack{A^T,A'^T}\right\rbrack^T =-\lbrack{A,A'}\rbrack$ and thus
\begin{equation}\label{BTa}
\bigl((B^{T_\alpha})^2\bigr)^{T_\alpha}
   = \sum_{\beta\rm\ even}\ (-1)^{\vert\alpha\cap\beta\vert}
         \bigotimes_{k\in \beta}{i\over2}\lbrack{A_k,A_k'}\rbrack.
\end{equation}
Note that it does not matter whether we transpose $\alpha$ or its
complement.

Now consider a partition of $\{1,\ldots,n\}$ into $p$ nonempty and
disjoint subsets $\alpha_1,\ldots,\alpha_p$. Let us denote by
${\cal P}$ the collection of all unions of these basic sets
together with the empty set, so that ${\cal P}$ has $2^p$
elements. We assume that $\rho^{T_\alpha}\geq0$ for all
$\alpha\in{\cal P}$. For $p=1$ this is no constraint at all,
because the full transpose of $\rho$ is always positive. At the
other extreme, for $p=n$, this assumption means the positivity of
every partial transpose.

We now take the expectation value of Equation~(\ref{BTa}), and
average over the $2^p$ resulting terms. The coefficient of the
$\beta^{\rm th}$ term then becomes
\begin{equation}\label{fourierP}
  2^{-p}\sum_{\alpha\in{\cal P}} (-1)^{\vert\alpha\cap\beta\vert}
  =2^{-p}\prod_{m=1}^p(1+(-1)^{\vert\alpha_m\cap\beta\vert} ).
\end{equation}
 which is proved by writing the sum over ${\cal P}$ as a sum over
$p$ two-valued variables, labeling the alternative
``$\alpha_m\subset\alpha$ or $\alpha_m\not\subset\alpha$'', and
using that the parity $(-1)^{\vert\alpha\cap\beta\vert}$ is the
product of the parities corresponding to the $\alpha_m$. Clearly,
the expression~(\ref{fourierP}) is one iff
${\vert\alpha_m\cap\beta\vert}$ is even for all $m$ and zero
otherwise. Let us call such sets $\beta$ ``${\cal P}$-even''.
There are
\begin{equation}\label{countPeven}
  \prod_m 2^{\vert\alpha_m\vert-1}=2^{n-p}
\end{equation}
 such sets. Hence we get the bound
\begin{eqnarray}
  ({\rm tr}\rho B)^2&\leq& \sum_{\beta\ {\cal P}{\rm-even}}
      {\rm tr}\Bigl(\rho\ \bigotimes_{k\in \beta}{i\over2}\lbrack{A_k,A_k'}\rbrack\Bigr)
  \nonumber\\
  &\leq&2^{n-p}.\label{bigbound}
\end{eqnarray}
 That this bound is optimal is evident by evaluating it on a
tensor product of pure states maximally violating Mermin's
inequality for each partition element $\alpha_m$, i.e., states as
discussed in Section~III.

To summarize, we have established the best bound
\begin{equation}\label{finalbound}
   \vert{\rm tr}(\rho B)\vert \leq 2^{(n-p)/2}
 \end{equation}
on violations of Mermin's inequalities, under the assumption that
the partial transposes $\rho^{T_\alpha}$ are positive for all
$\alpha\subset\{1,\ldots,n\}$ subordinated to a partition into $p$
subsets. This includes three special cases: For $p=1$ it is the
analogue of Cirel'son's inequality, for $p=n$ it proves our claim
that the inequalities are satisfied if all partial transposes are
positive, and for partitions of the form
$\{1\},\ldots,\{m\},\{m+1,\ldots n\}$, we obtain the result by
Gisin et al.\cite{Gisin} using Mermin's inequalities to test for
the number $m$ of independent qubits.

\end{document}